\documentclass[doublecol]{epl2}
\usepackage{epsfig}

\usepackage{graphicx}% Include figure files
\usepackage{dcolumn}% Align table columns on decimal point
\usepackage{bm}% bold math
\usepackage{epstopdf}
\newcommand{\ac}{{\rm ac}}
\newcommand{\bd}{{\rm bd}}
\newcommand{\jc}{{\rm jc}}

\newcommand{\yr}{{\rm yr}}

\newcommand{\vb}{{\sigma_\jc}}

\newcommand{\km}{{\rm km}}

\newcommand{\mBD}{$\mu$BD}
\newcommand{\mBDs}{$\mu$BDs}

\newcommand{\bu}{{\rm bu}}
\newcommand{\gc}{{\rm gc}}

\renewcommand{\sf}{{\rm sf}}

\newcommand{\bh}{{\rm bh}}
\newcommand{\si}{{\rm s}^{-1}}

\newcommand{\apj}{apj}

\newcommand{\half}{{\frac{1}{2}}}

\newcommand{\myskip}{\footnote}%[1]{}
\renewcommand{\myskip}[1]{}
\newcommand{\emphf}[1]{}

\renewcommand{\d}{{\rm d}}

\newcommand{\gal}{{\rm gal}}

\newcommand{\BEQ}{\begin{eqnarray}}
\newcommand{\EEQ}{\end{eqnarray}}
\newcommand{\BEA}{\begin{eqnarray}}
\newcommand{\EEA}{\end{eqnarray}}

\newcommand{\m}{{\rm m}}

\begin{document}

\title{Model for common growth of supermassive black holes, bulges and globular star clusters: Ripping off  Jeans clusters}

\shorttitle{Model for common growth of supermassive black holes, bulges and globular star clusters\date{today}}

\author{Theo M. Nieuwenhuizen$^{1,2}$}
\shortauthor{Th. M. Nieuwenhuizen}

\institute{$^{1}$ Institute for Theoretical Physics, University of Amsterdam, Science Park 904, P.O. Box 94485,  1090 GL  Amsterdam, The Netherlands
 \\ $^{2}$ Center for Cosmology and Particle Physics, New York University,  4 Washington Place, New York, NY 10003, USA }

%\pacs{98.80.Bp}{Origin and formation of the universe}



% \pacs{95.36.+x}{Dark energy}

\abstract{ %\begin{abstract}
It is assumed that a galaxy starts as a dark halo of a  few million Jeans clusters (JCs), each of which consists of nearly a trillion micro brown dwarfs, MACHOs
of Earth mass. JCs in the galaxy center heat up their MACHOs by tidal forces, which makes them expand, so that coagulation and star formation occurs.
Being continuously fed by matter from bypassing JCs, the central star(s) may transform into  a super massive black hole. 
It has a fast $t^3$ growth during the first mega years, and a slow $t^{1/3}$ growth at giga years.
JCs disrupted by a close encounter  with this black hole can provide matter for the bulge.
Those that survive can be so agitated that they form stars inside them and become globular star clusters.
Thus black holes mostly arise together with galactic bulges in their own environment and are about as old as  the  oldest globular clusters.
The age 13.2 Gyr of the star HE 1523-0901 puts forward that the Galactic halo was  sufficiently assembled at that moment.
The star formation rate has a maximum at black hole mass $\sim4 \ 10^7M_\odot$ and bulge mass $\sim5\,10^{10}M_\odot$.
In case of merging supermassive black holes the JCs passing near the galactic center provide ideal assistance to overcome the last parsec.}
%\end{abstract}

\maketitle

\email{$^{(a)}$t.m.nieuwenhuizen@uva.nl\date{today}}

\section{Introduction}
How galactic bulges and their central, supermassive black holes (BHs) have formed is still a mystery. Where did the matter come from in the first place?
Did they grow  slowly in their own environment or mainly by merging? How is their growth related to that of the bulge?
Why do the heaviest active galactic nuclei involve about ten billion solar masses?
With new evidence coming in steadily, these questions are now often debated but still await consensus, see the overview by  \cite{Heckman2011}.

Recent observations reveal  that supermassive black holes (SMBHs) of up to
 billions of solar masses must have formed rather early in the history of the universe and 
 that this happened rather smoothly, with black holes growing in tandem with their hosts throughout cosmic history, starting from the earliest times  \cite{Treister2011}.
It is seen that most copiously accreting black holes at these epochs are buried in significant amounts of gas and dust that absorb most radiation 
except for the highest energy X-rays. This suggests that black holes grow significantly more than previously thought during these early bursts.
In line with this, it is observed that merging can not be the driving force for BH growth~\cite{Kocevski2011}.
On the other hand, the BH mass exhibits no correlation with the dark matter in the halo \cite{Kormendy2011}.
 
There is also evidence for correlations between galactic structures: the mass of the central supermassive black hole (SMBH) correlates with the mass 
of the bulge \cite{HaringRix2004} and also with the number of globular star clusters \cite{BurkertTremaine2010}. 
These intriguing relations point at a common dynamical origin.

The standard cold dark matter (CDM) paradigm fits the cosmic microwave background data  fairly
well\footnote{WMAP has too many $2-\sigma$ points. The best fit, $\chi^2/\nu=1.06$ for $\nu\sim1000$ degrees of freedom,
exceeds the Gaussian $\chi^2=\nu$ and leaves space for a
 $\sqrt{60}\sigma\approx8\sigma$ improvement~\cite{Nolta}, 
a conundrum which the Planck data will not resolve~\cite{LymanPage}.},
but there are ongoing debates about the analysis of WMAP data, see e.g. \cite{Shanks2011}. Another weakness is
that correlations in galaxy structures at the largest observed scales are stronger than expected from $\Lambda$CDM ~\cite{Sylos2011}.
The CDM particle has still not been found, and time for it is running out~\cite{Bertone2010}.
At the time of this writing, December 2011, no trace of supersymmetry has been observed at the Large Hadron Collider LHC,
which dims the hope to ever observe the CDM particle there.
Worse, there are arguments against a CDM particle, since CDM performs badly at the galactic scale. 
Most known is the missing satellite problem,  from the prediction of numerous satellites of the Galaxy. Not only are they not observed,
it is now becoming  generally accepted that the known satellites lie in a plane,
not in random CDM subhalos, and have come in from the same direction, 
possibly in a collision between the Milky Way and the progenitor of the Magellanic Clouds \cite{Pawlowski2011}.
On top of this, low surface brightness (dwarf) galaxies contain a lot of dark matter and should be an ideal testing ground,
but CDM fails badly in describing them~\cite{Kroupa2010}.
Scientists like C. Frenk (of Navarro-Frenk-White-profile fame) now investigate other scenarios \cite{FrenkWDM2011},  
a fact that has reached headlines\footnote{http:$//$www.bbc.co.uk/news/science-environment-14948730},
while also J. Navarro concludes that dark matter halos of dwarf galaxies pose a challenge for the $\Lambda$CDM  paradigm
\cite{FerreroNavarro2011}.
These studies break with the attitude within CDM circles  to see these phenomena as ``(dirty) gastrophysics'',
issues that need not be addressed at first.

The merit of MOND (MOdified Newtonian Dynamics) is to establish a protocol
to predict in galaxies the force, and thus the circular rotation speed, from the luminous matter and gas alone. 
This does not prove that MOND is right -- we are not convinced of it  -- but it concludes that 
{\it cold dark matter cannot explain galactic rotation curves}. Indeed, being non-dissipational 
it has no clear reason to act in the same way as the dissipative baryons~\cite{Sanders}.

These problems of CDM motivate us to investigate a different starting point, namely
the old case of Galactic dark matter dominated by Earth mass MACHOs.
Such objects were observed in quasar microlensing by \cite{Schild1996} 
and in later works on quasars ~\cite{Pelt1998}, 
a result that has not been disputed and is supported by studies on other quasars.
On the other hand, the EROS, MACHO and OGLE consortia did not 
observe them in front of the Magellanic Clouds, with the relevant papers for Earth masses being
from EROS \cite{Renault1998} and from MACHO \cite{Alcock1998,Alcock2000}.
Despite the quasar detection claim, their non-existence is now the prevailing opinion.
But there is mounting evidence that the low-mass MACHO observations have lacked the required precision.
The MOA team recently found a new population of Jupiters in the Galactic bulge, twice as common as main sequence stars
\cite{Sumi2011}. MOA-II announced new high cadence observations toward Galactic bulge and Magellanic clouds: 
Until now,  11 extrasolar planets, 10 free-floating planets, and four MACHO candidates have been discovered \cite{Abe2011}.

To resolve the conundrum of MACHO observation versus non-observation,  it is planned to redo the
MACHO searches in front of the Small Magellanic Cloud with the large 6.5 m telescope in Chili \cite{Protopapas2011}.

We thus consider it appropriate to assume MACHO dark matter.
In this letter, we recall its gravitational hydrodynamics basis and 
apply that to the growth of BHs and to bulges and globular star clusters.
Then we consider  the SMBH at Sag A$^\ast$ and we close with a discussion.

\section{Elements of Gravitational hydrodynamics} \noindent\noindent
We start from the well accepted fact that soon after the decoupling of matter and radiation the newly formed neutral gas breaks up in
Jeans clumps of some 600,000 solar masses. From gravitational hydrodynamics (GHD) we take one more
ingredient: due to turbulence and viscosity, the Jeans clumps are supposed to  fragment into hydrogen-helium  
 balls of Earth mass~\cite {Gibson1996,NGS2009}, now termed micro brown dwarfs (\mBDs) or MACHOs. 
Thereby  Jeans
 clumps become Jeans clusters (JCs) of some $2\, 10^{11}$ \mBDs~\cite{NSG2011}.

The \mBDs \  are primordial objects basically consisting of H and 26\% in weight of $^4$He.  
Visible matter, such as stars and hydrogen clouds, should have arisen
from them, but most \mBDs \  should still exist, having picked up a variety of metals from their
environment. In this picture the dark matter halo of the  Galaxy is made up of some $10^6$ Jeans clusters.
Essentially, they are the progenitors of globular star clusters.
If their joint distribution is an isothermal sphere, the flattening of rotation curves is explained.

This starting point has already explained many puzzling features, such as the formation of 
young globular clusters in galaxy merging and the iron planet-core problem \cite{NSG2011}.
 It also explains the Helium-3 problem \cite{NHelium32011}.
Some Galactic phenomena are so frequent that they are likely related to DM.
First, there is the ubiquitous minimal 15 K temperature of  ``cold cirrus dust''
\cite{Veneziani2010}. Second, there occur ``long duration radio transients'', mysterious radio events that last
long (longer than half an hour but shorter than a few days), are very frequent 
($\sim1000$ deg$^{-2}$yr$^{-1}$, i.e., a new one every second if the rate holds over the $4\pi$ sky), 
can be very bright ($>1$ Jy), have neither a precursor nor a remnant, and are not visible in the infrared,
visible or X-ray spectrum \cite{Ofek2010}. Both phenomena can be explained by DM from Earth mass MACHOs
\cite{NSG2011} but have  little chance of explanation by any non-baryonic DM, emergent--gravity DM \cite{Verlinde2011}, 
or by modifications of Newton's law such as MOND.

The typical size of an \mBD \ can actually be estimated from isothermal modeling (see the discussion below),
$R_\bd=GM_\bd/2\sigma_{\rm gas}^2$. Here $\sigma_{\rm gas}=(k_BT/\mu m_N)^{1/2}\simeq450$ km/s, using
the observed minimal temperature $T\simeq15 $ K  of what is commonly called ``cirrus clouds'' but expected by us to be \mBDs \
grouped  in JCs~\cite{NSG2011}; furthermore, $\mu\simeq 0.61$ is the 30\% solar metallicity factor and $m_N$ the nucleon mass.
For $M_\bd=M_\oplus$ it follows that $R_\bd\simeq 9.8 \ 10^5$ km $=1.4 \ R_{\odot}$, which agrees surprisingly well with the 3.7 $R_\odot$ 
from the typical duration $\sim 1$ day of  long duration radio transients as explained by merging of two \mBDs \ with 
typical speeds of 30 km/s~\cite{NSG2011}. Though it may seem that an Earth mass  of gas smeared over a solar size is 
dilute, the surface density of H atoms, $0.74M_\bd/m_N\pi R_\bd^2\sim 3\ 10^{28}{\rm cm}^{-2}$, is very large for a cosmic cloud. 
That the \mBDs \ have so far escaped direct detection is explained by the fact that their individual size and the fraction of the sky 
they cover together are small~\cite{NSG2011}, and that below 20 K they are liquid.
%in the liquid phase.

We may also mention Segue 1, the darkest object known with about 1000 stars and a mass-to-light ratio of 3400. 
Its total mass within the half-light radius is 600,000 solar masses \cite{Simon2011}.
Clearly Segue 1 is DM dominated but no X-ray signal has been observed, so no annihilation or decay of CDM.
  The GHD explanation is simple: Segue 1 is a Jeans cluster with little star formation and its DM is baryonic.
The same holds for similar low brightness star clusters.

Non-baryonic dark matter must exist too and may consist of non-relativisitic neutrinos that appear as the dark matter of clusters \cite{Nnu2009}.
The predicted masses of 1.5 eV (if all nearly equal), or in the eV range in general,
will be tested in the KATRIN experiment  in 2014 \cite{NieuwenhuizenMorandi2011}.
If confirmed, the two-types-of-DM combination of MACHO dark matter in galaxies and neutrino dark matter in clusters 
can make cold dark matter obsolete \cite{Nnu2009}.

\section{Growth of Supermassive BHs} {\it The physical picture}. 
The present work aims to point out a dynamical connection between 
JCs, galactic bulges and central BHs.
The basic idea is that a galaxy starts as a dynamically bound  halo of dark  Jeans clusters 
-- stars did not form. 

To analyze the ``best scenario'' we shall work out the case where the JCs have a singular isothermal distribution.
In the center the density will be large enough to first trigger planet formation from collisions between 
the \mBDs \  and later the formation of one or more stars and then a central black hole.
The constant stream of by passing JCs will provide the opportunity for the BH to 
catch \mBDs, thus material to grow. When the BH is large enough, exceeding $3\ 10^8M_\odot$,
it can catch complete JCs. This mechanism explains central BH growth in the absence of galaxy merging.
JCs that are disrupted by the BH may change into giant molecular clouds or form the stars in the bulge, while JCs that are 
only agitated may survive but develop star formation and turn into globular star clusters.
So all these phenomena are related, they are fed by the dark JC halo and correlated with it.

\subsection{The seeding phase}

The singular isothermal distribution of matter,
\BEQ \label{rhoiso}
\rho=\frac{\sigma_\jc^2}{2\pi G r^2},
\EEQ
has included mass ${\cal M}(r)=2\sigma_\jc^2r/G$. 
We assume that (\ref{rhoiso}) holds for the distribution of JCs that build the galactic halo,
and we take for the velocity dispersion $\sigma_\jc=200$ km/s.
This relates the galactic mass and radius,

\BEQ 
M_\gal=\frac{2\sigma_\jc^2}{ G }R_\gal.
\EEQ
A similar shape for \mBDs \ inside a JC with $\sigma_\bd= 30$ km/s leads to

\BEQ 
M_\jc=\frac{2\sigma_\bd^2}{ G }R_\jc. \qquad
\EEQ 
In the center of the galaxy a number of JCs overlap, 

\BEQ 
\label{Sigma=}
\Sigma^2\equiv \frac{{\cal M}(R_\jc)}{M_\jc}=\frac{\sigma_\jc^2}{\sigma_\bd^2}=44. 
\EEQ
Hence a singular core has a crowded center with several JCs crossing each other. Tidal forces will heat the \mBDs,
which makes them expand, so that they may coagulate. 
Thus planet and star formation out of \mBDs \ is likely to happen and hence the formation of a central BH.

The Newton force at a distance $r$ from the BH is $GM_\bh/r^2$ and the statistical one of JCs
is $G{\cal M}(r)/r^2$. Equating them defines the ``active'' radius $R_\ac$ through ${\cal M}(R_\ac)=M_\bh$
~\cite{BinneyTremaine},
inside which the BH strongly modifies the dynamics of the JCs and their \mBDs.
This provides a feeding mechanism for the BH.

What is the minimal central mass that can disturb the bypassing \mBDs? 
Their typical \mBD \ distance in the $\Sigma^2$ overlapping JCs is $\ell_\bd=R_\jc (M_\bd/M_\jc\Sigma^2)^{1/3}$. 
A sphere of this radius contains typically one  \mBD, so
every central object, be it  a planet, a star or a BH, has $R_\ac>\ell_\bd$, so it will disturb the by passing \mBDs, 
 presenting a natural mechanism for them to grown from  the material in the \mBDs.

\subsection{Initial growth phase}

At a scale $R_\ac\ll R_\jc$ in the center we deal with a uniform distribution of the matter of (\ref {Sigma=}), 

\BEQ {\cal M}(R_\ac)=\Sigma^2 M_\jc \frac{R_\ac^3}{R_\jc^3}.
\EEQ
The number of \mBDs \ that enter a sphere of radius $R_\ac$ between  $t$ and $t+\d t$ is
(we leave out factors of order unity) 

\BEQ \d N_\bd= \Sigma^2 \frac{M_\jc}{M_\bd} \frac{R_\ac^2\sigma_\jc\d t}{R_\jc^3}.
\EEQ
Our assumption is that the BH accretes a fraction $f_\bh^i={\cal O}(1)$ of the mass involved in this,

\BEQ \dot M_\bh =f_\bh^i \dot N_\bd M_\bd\equiv\frac{3}{\tau}M_\bh^{2/3}M_\odot^{1/3},
\EEQ
which defines  the characteristic time scale

\BEQ
\tau=\frac{3R_\jc}{f_\bh^i \sigma_\jc}\, \frac{M_\odot^{1/3}}{\Sigma^{2/3}M_\jc^{1/3}}=\frac{1}{f_\bh^i}650\, {\rm yr}.
\EEQ
The solution of this dynamics,

\BEQ \label{Mbhshort}
M_\bh=M_\odot\left(\frac{t}{\tau}\right)^3,
\EEQ
marks an explosive growth, producing a star in a timescale of a thousand years, which transforms into a SMBH
by ``eating''  too much \mBDs. It will grow up to a $10^6M_\odot$ SMBH in the rather short time of circa 0.1 Myr.

\subsection{Final growth phase}

The growth mechanism is the more complicated the heavier the BH, in particular for $M_\bh\sim10^6M_\odot$.
However, it becomes simpler for the supermassive ones, where a whole JC can be captured. 
In this regime the relation ${\cal M}(R_\ac)=M_\bh$ yields

\BEQ \label{Raciso}
R_\ac=\frac{GM_\bh}{2\sigma_\jc^2}.
\EEQ
which exceeds $R_\jc$ when $M_\bh {>\atop\sim} M_\bh^\ast\equiv\Sigma^2M_\jc=2.7\,10^8M_\odot$.
The JCs entering the sphere of radius $R_\ac$ in a time interval  $\d t$ were located between 
$R_\ac$ and $R_\ac+\sigma_\jc\d t$ and had $\dot r<0$, so the average rate of JCs entering  per unit of time is 

\vspace{-3mm}

\BEQ \label{Ndotcusp1}
 \dot N(R_\ac)=\frac{\rho(R_\ac)}{M_\jc}2 \pi R_\ac^2 \vb= \frac{\sigma_\jc^3 }{GM_\jc}=\frac{\Sigma^2\vb }{2R_\jc}.
 \EEQ
It is constant because the surface factor $R_\ac^2$ cancels the decay of  (\ref{rhoiso}).
The probability for a JC to cross the central BH is set by the opening angle \cite{BinneyTremaine},

\vspace{-3mm}

\BEQ
\label{coneprob1}
\frac{\pi R_\jc^2}{2\pi R_\ac^2}=\half \left(\frac{R_\jc\sigma_\jc^2}{GM_\bh} \right)^2=\half\left(\frac{M_\jc\Sigma^2}{M_\bh} \right)^2.
\EEQ
A fraction $f_\bh^f={\cal O}(1)$ of the JC mass is supposed to end up in the BH and the rest in the bulge or back in the halo.
Putting (\ref{Ndotcusp1}) and (\ref{coneprob1})  together, we get

\vspace{-3mm}

\BEQ
\dot M_\bh=\frac{M_\jc^3}{3\tau_\ast M_\bh^2},
\EEQ
with 

\vspace{-3mm}

\BEQ 
\tau_\ast=\frac{2GM_\jc}{3f_\bh^f \Sigma^4\sigma_\jc^3}=\frac{0.106}{f_\bh^f}\, \yr.
\EEQ
The solution is 

\vspace{-3mm}

\BEQ \label{Mbhlong}
M_\bh=M_\jc\left(\frac{t}{\tau_\ast}\right)^{1/3}. 
\EEQ
This behavior holds for $M_\bh\gg 2.7\,10^8M_\odot$. 
While (\ref{Mbhshort}) exhibits a very fast growth at early times, Eq. (\ref{Mbhlong}) shows that this is strongly slowed down at late times.
The result takes the values $M_\bh=(1,2,3,18)\,10^9M_\odot$ at times  $t=(0.51, 4.2, 14, 3000)(1/f_\bh^f)$ Gyr. 
So though masses of a few times $10^8M_\odot$ should be quite common,  no mass is predicted beyond  a few billion solar masses.
Merging could stretch this upper limit somewhat; the most massive one contains 18 billion solar masses \cite{Valtonen2008}.
Indeed, in the nearby universe unexpectedly heavy SMBHs have been observed recently. 
In NGC 3842, the brightest galaxy in a cluster at a distance from Earth of 98 megaparsecs, the BH has a central black hole with a mass of 
9.7 billion solar masses; A black hole of comparable or greater mass is present in NGC 4889, the brightest galaxy in the Coma cluster,
which lies at a distance of 103 megaparsecs~\cite{McConnell2011}.

The age at redshift $z$ in a cosmology with matter fraction $\Omega_M=0.3$ and Hubble constant $H_0=72$ km/s Mpc is 
 $t(z)={2}/{3H_0\sqrt{\Omega_M}(z+1)^{3/2} }$ at redshift $z>1$,
so that the maximum BH mass at large redshift is

\vspace{-3mm}

\BEQ  M_\bh^{\rm max}(z)=\frac{3.7\cdot10^9}{\sqrt{z+1}}M_\odot.
\EEQ
This typical $z$-dependence can be tested on  high-$z$ SMBHs.

\section{Bulge growth and giant molecular clouds}

Not all the material of JCs that come close to the BH will end up in it. 
Heavily distorted JCs will start to create the bulge. 
Some of them can heat up enough to make most of the \mBDs \ dissolve,  
turning the remnant of the JC into a giant molecular cloud -- thus explaining their origin.
Indeed, giant molecular clouds can have a mass of 
100,000 -- 400,000 
% 1--4 $10^5$ 
times the solar mass, while
our canonical value for the mass of a JC is 
% 6 $10^5$ $M_\odot$.
 600,000 $M_\odot$.
In these clouds star formation can occur, in particular if there are still nuclei of original \mBDs, or intact ones,
that can aggregate to form stars~\cite{SchildGibson2011,GibsonSchild2011}.

In the simplest model  rate of increase of bulge mass will be proportional to the BH mass growth rate,

\vspace{-3mm}

\BEQ \dot M_\bh=f_\bu \dot M_\bu
\EEQ
and if $f_\bu$ can be taken as constant, its integral will be

\vspace{-3mm}

\BEQ  \label{MbhMbu}
M_\bh=f_\bu  M_\bu
\EEQ
Observations show that \cite{HaringRix2004} 

\vspace{-3mm}

\BEQ 
M_\bh=(1.4\pm0.4)10^{-3}M_\bu \quad  \textrm{at} \quad M_\bu=5\, 10^{10}M_\odot,
\EEQ
so that  $ f_\bu= 0.0014$, while  Ref. \cite {Heckman2011} gives the typical value $ f_\bu\sim 0.001$.
Ref. \cite{Cisternas2011} reports that $f_\bu$ has remained constant in the last 7 billion years (out to $z=0.9$).

In a linear modeling  the star formation rate (SFR) will also be proportional to the BH growth,

\vspace{-3mm}

\BEQ 
 \dot M_\bh= f_{\rm sf} \ {\rm SFR}.
\EEQ
The relationship between black hole growth and star formation is investigated in Seyfert galaxies \cite{DiamondStanicRieke}.
The authors study masses  between $3\cdot 10^5$ and $6\cdot10^8M_\odot$ and deduce the star formation rate from 
near-infrared observations at 
$1.13\ \mu \m$ 

\vspace{-3mm}

\BEA
{\rm SFR}(1.13\ \mu \m)=14^{+11}_{-6}\left( \frac{\dot M_\bh} {M_\odot\, {\rm yr}^{-1}}\right)^{0.95\pm0.10}\frac{M_\odot}  {{\rm yr}} .
\EEA
Taking this power equal to unity,  we observe that this fits within our picture, and we get the estimate
$f_{\rm sf}\sim 0.07$. 

We notice a discrepancy between the amount  of mass entering the bulge $\sim f_\bu^{-1}\dot M_\bh$  and the
mass in star formation  $\sim f_\sf^{-1} \dot M_\bh$, which may imply that the factors $f_\bu$ and $f_\sf$ are not constants.

The SFR equals $f_\sf^{-1}\dot M_\bh$, which grows for small BH mass as $M_\bh^{2/3}$ according to Eq. (\ref{Mbhshort}),
while it decays as $M_\bh^{-2}$ according to (\ref{Mbhlong}). If $f_\bh^i\approx f_\bh^f$ these asymptotic behaviors cross
at $M_\bh^\ast=3.7 \ 10^7M_\odot$ and, with (\ref{MbhMbu}),  $M_\bu^\ast\sim 5\, 10^{10}M_\odot$.
This corresponds to %The crossing of these asymptotes occurs at 
SFR  $=500f_\bh^f M_\odot$yr$^{-1}$, a reasonable upper bound for the maximal SFR.

\section{ Globular star cluster formation} 

It is commonly assumed that globular star clusters arise through the Jeans mechanism.
The combination of the fragmented structure of JCs and the agitation by the SMBH provides
a mechanism to induce star formation and hence transform JCs into globular clusters.  
Some JCs will pass by close enough to the BH to be agitated by tidal forces, though remaining 
enough intact to go back into the halo. When the \mBDs \ get heated so that they expand and coalesce,
they form stars. This may in the end yield a number of globular clusters (GCs)  proportional
to the BH mass,

\vspace{-3mm}

\BEQ N_\gc=f_\gc\frac{M_\bh}{M_\jc}.
\EEQ
Burkert and Tremaine \cite{BurkertTremaine2010} were the first to investigate a possible connection
between the number of globular clusters and they  BH mass. They derive from observations

\vspace{-3mm}

\BEQ
 \label{MbhNgc}
\frac{M_\bh}{M_\odot}=1.7\cdot10^5N_\gc^{1.08\pm0.04},\qquad \frac{M_\bh}{M_\jc}=0.283 \ N_\gc,
\EEQ
yielding $f_\gc\sim 3.5$. Forcing the slope to be 1, the best fit is~\cite{HarrisHarris}

\vspace{-5mm}

\BEQ 
\frac{M_\bh}{M_\odot}=4.07 \ 10^5N_\gc,\qquad 
\EEQ
so that we get the estimate $ f_\gc\sim1.5$.

Such a transformation of dark JCs into young GCs  is believed to happen also
in galaxy mergers, where young globular star clusters arise long after the merging process
has taken place ~\cite{Gibson1996,NSG2011}. An example is the Tadpole
galaxy, where about 11,000 GCs have been analyzed by~\cite{Fall2005}.
The most luminous of the ``knots'' have an age of 4--5 Myr and estimated mass  
$6.6\cdot10^5M_\odot$~\cite{Tran2003}, reminiscent of a JC.

\section{Solution to the last parsec problem} 
It has long been suspected that SMBHs arise from merging of smaller ones. 
Though we have proposed a different main mechanism, merging will definitely also occur. 
To merge, a BH pair can scatter a star and become more tightly bound. 
But the dynamical friction with the stellar background is ineffective in shrinking the binary 
below separations of  1 parsec \cite{Begelman1980,MilosavljevicMerritt2001}.
This conundrum has puzzled the community for decades, see, e.g., \cite{Khan2011}.
But GHD offers a simple way out: galactic centers are crowded with JCs of \mBDs.
The JC size is in the parsec regime, so they offer an ideal frictional environment for a 
rapid merging of the BHs.

\section{The Galaxy and its SMBH at Sag A$^\ast$}
The Sun is located at 8 kpc from the Center of the Galaxy. It is a cored Sersic galaxy with a bulge surface density 
$\Sigma_\ast=\Sigma_0\exp(-r/R_d)$ with $R_d\approx 2.5$ kpc.
In the Center there is a nuclear star cluster and  the density has a cusp, with $M_\ast\sim 10^7M_\odot$ at $r<4$ pc
\cite{DoGhez2009}.
These aspects stem with our picture of central JCs. 
Our estimate for the mass of singular isothermal cores, ${\cal M}(R_\jc)=2\Sigma^2 M_\jc$ yields the right order of magnitude,
$5.3\ 10^7M_\odot$ for $R_\jc=1.43$ pc.

The central BH of our Galaxy appears to verify several fundamental aspects of our picture.
Located at Sag A$^\ast$ it has mass of $4\ 10^6M_\odot$, a  modest value for a supermassive BH.
Observations have revealed a puzzling disk of over $50$ young stars (age $\sim$ 1 Myr) within 0.14 pc of Sag A$^\ast$ 
that probably formed in situ but in a more complex geometry than a simple, thin circular disk.
Lacking a clear explanation, this conundrum has been termed the  ``paradox of youth" ~\cite{LuGhez2009}.
A new scenario is offered by GHD: a Jeans cluster was passing close to  Sag A$^\ast$. 
This has agitated the \mBDs \  by tidal forces, so that they have grown in size, merged and finally turned into stars,
in the same way as they do in galaxy mergers~\cite{Gibson1996,NSG2011}.
Indeed, the active radius (\ref{Raciso})  takes the value $R_\ac=0.2$ pc, which is comparable with the observed 0.14 pc radius.
This view is supported by the fact that the observed out-of-the-disk velocity dispersion of the young stars of $28\pm  6\  \km\ \si$~\cite{LuGhez2009}
fits well with the typical  $\sigma_\bd= 30$ km $\si$ velocity dispersion of  \mBDs \ inside JCs.

The scenario of by passing JCs is supported by the fact that the
 very old globular cluster NGC 6522 is (presently) situated close to the Galactic nucleus \cite{vdBergh2011}.
Also in M31 (Andromeda) a cluster of blue (young) stars is found to surround the nuclear black hole; its size, estimated at a few parsec,
corresponds to a Jeans cluster, as we would expect.
Like the black hole in the Milky Way, the one in M31 is closely surrounded by apparently young stars~\cite{Lauer2011}.

\section{Conclusion}
The prediction of gravitational hydrodynamics that after the decoupling the newly formed Jeans gas clumps fragment in micro brown dwarfs
of Earth weight is considered here under the assumption that these Jeans clusters (JCs) build the galactic halo by reaching a dynamical quasi equilibrium  
that we model by a singular isothermal sphere.
This picture provides a simple answer for the growth of central (supermassive) black holes, the bulge, giant molecular clouds and globular star clusters.
The mass of the BH grows quickly, $\sim t^3$, at early times, and a $10^6M_\odot$ mass can be accumulated in, say, 100,000 yr.
The final growth is slower, $\sim t^{1/3}$, and in a Hubble time a black hole of weight up to a few billion solar masses can grow. 
These different behaviors imply a maximal star formation rate at BH mass of $\sim 4\cdot10^7M_\odot$ and bulge mass of $\sim 5\, 10^{10}M_\odot$.
Merging may create even heavier BHs because the last parsec is again overcome with help of Jeans clusters at the galactic center.
The heaviest SMBHs  (as well as the typical ones) are expected to have a mass scaling as $1/\sqrt{1+z}$ for $z> 1$.

In our top-down approach the observed age 13.2 Gyr of the star HE 1523-0901 \cite{Frebel2007} puts forward that the Galactic halo was 
sufficiently assembled at that moment.

In the present work it has been tacitly assumed that the isothermal distribution remains singular at the galactic center and exhibits no depletion. Such an effect 
is likely to occur, however, leading to a pause between spurts of black hole growth, as observed \cite{Trakhtenbrot2011}
and often connected to BH jets. This may explain both why the Galactic BH at Sag A$^\ast$ is 
quiescent  and why some black holes are quasars and others not. The repletion mechanism may also address the Faber-Jackson relation and variants of it;
they are not explained in our approach, but repletion may imply a common cause for them.

We have presented a basic model for these behaviors.
Many details can be learned from numerical simulations.

{\it Additional remark:} After submitting  the manuscript we noticed surprising observations that
support our theory but are unexpected from $\Lambda$CDM.  
1) A cloud of $3M_\oplus$ heads towards the supermassive BH at Sag A$^\ast$, the center of our Galaxy \cite{Gillessen2011}. 
We view it as a typical \mBD \  feeding the BH, after disruption from a by-passing JC. 
The many \mBDs \ of the JC can absorb angular momentum to facilitate the cloud's central orbit. 
 A test is to identify this JC and related gas clouds.  
 2)  Ref. \cite{Fumagalli2011} reports two pristine clouds without metals at $z=3$.
The smaller one has H-mass $4.2\,10^5M_\odot$, as expected for a nearly standard JC of $5.7\ 10^5M_\odot$ in H and He.
 We explain them as JCs that were long outside the bulge, thus having picked up few metals.
Upon heating,  the \mBDs \ evaporated and the JCs expanded.
3) In accordance with the GHD top-down structure formation,
the higher the stellar mass, the lower the age of the Universe at which certain $z>4.7$ galaxies formed \cite{Castro2011}.
4) Ultra-compact dwarf galaxies are consistent with just being the bright tail of the 
globular cluster population (JCs with star formation) rather than being tidally transformed  dwarf galaxies \cite{Mieske2012}. 
5) A large star formation rate ($\sim100 \,M_\odot\, {\rm yr}^{-1}$) occurs already in a very early
galaxy at 12.9 Glyr distance ($z=7.2$) \cite{Ono2012}. Our theory allows this, and connects it a growing 
supermassive black hole.
%SMBH.

\myskip{
We quote from %the NASA press release 
\cite{Nasa2011}:
 ``Astronomers using NASA's Spitzer and Hubble space telescopes have discovered that one of the most distant galaxies known is churning 
out stars at a shockingly high rate. The blob-shaped galaxy, called GN-108036, is the brightest galaxy found to date at such great distances.

The galaxy, which was discovered and confirmed using ground-based telescopes, is 12.9 billion light-years away. Data from Spitzer and Hubble were used to measure the galaxy's high star production rate, equivalent to about 100 suns per year. For reference, our Milky Way galaxy is about five times larger and 100 times more massive than GN-108036, but makes roughly 30 times fewer stars per year. 

The discovery is surprising because previous surveys had not found galaxies this bright so early in the history of the universe.'' This observation fits
in our theory, and we expect a SMBH to grow in its center. The central 
}

\end{document}